# Landau quantization and the thickness limit of topological insulator thin films of Sb$_2$Te$_3$

Yeping Jiang[1,2], Yilin Wang[1], Mu Chen[1,2], Zhi Li[1], Canli Song[1,2], Ke He[1], Lili Wang[1], Xi Chen[2], Xucun Ma[1*], and Qi-Kun Xue[1,2*]

[1] *Institute of Physics, Chinese Academy of Sciences, Beijing 100190, People's Republic of China*

[2] *State Key Laboratory of Low-Dimensional Quantum Physics, Department of Physics, Tsinghua University, Beijing 100084, People's Republic of China*

We report the experimental observation of Landau quantization of molecular beam epitaxy grown Sb$_2$Te$_3$ thin films by a low-temperature scanning tunneling microscope. Different from all the reported systems, the Landau quantization in Sb$_2$Te$_3$ topological insulator is not sensitive to the intrinsic substitutional defects in the films. As a result, a nearly perfect linear energy dispersion of surface states as 2D massless Dirac fermion system is achieved. We demonstrate that 4 quintuple layers are the thickness limit for Sb$_2$Te$_3$ thin film being a 3D topological insulator. The mechanism of the Landau level broadening is discussed in terms of enhanced quasiparticle lifetime.

[*] Corresponding authors. Email: xcma@aphy.iphy.ac.cn, qkxue@mail.tsinghua.edu.cn

A topological insulator (TI) [1-2] is characterized by nontrivial gapless states on its surface/boundary due to strong spin-orbit coupling [3]. The surface states of a strong TI are massless Dirac fermions (DFs) and consist of an odd number of Dirac cones, centering at the time-reversal invariant momentums that are robust against any time-reversal invariant perturbation [4]. Recently, a class of strong TIs ($Bi_2Te_3$, $Bi_2Se_3$ and $Sb_2Te_3$) with a relatively large bulk energy gap and a single surface state Dirac cone has been theoretically predicted [5] and experimentally proved [6-8]. The helical surface states lead to a non-zero Berry's phase ($\pi$) of the electron wavefunction and a series exotic phenomena such as absence of back-scattering [9-12] and weak antilocalization [13]. In the case of thin films, at a critical thickness the surface states from opposite surfaces of the films can couple together and open a thickness-dependent gap [14-16], which is nontrivial and may give rise to quantum spin Hall state (QSH) similar to the case of HgTe quantum wells [17].

Despite extensive reports on TIs recently, the experimental study on $Sb_2Te_3$ is unexpectedly rare [8, 18]. The main reason is the material. Due to the relatively weak bonding between Sb and Te and the molecular nature of both Sb and Te beams, Sb vacancies and anti-site ($Sb_{Te}$) defects can easily form. Thus, as-grown $Sb_2Te_3$ is heavily p-doped with its Fermi level ($E_F$) lying in the bulk valence band. As shown in this study, the situation can be changed by growing high quality crystalline $Sb_2Te_3$ films by molecular beam epitaxy (MBE). The high quality films grown by MBE allow us to investigate topological surface states and the effect of those intrinsic defects directly by *in situ* scanning tunneling microscopy/spectroscopy (STM/STS). In spite of a large number of intrinsic substitutional defects, our STS measurements reveal well-defined Landau quantization in $Sb_2Te_3$ films, from which a nearly ideal linear dispersion of surface states can be deduced. By studying the thickness-dependence of surface states, we find that the thickness limit for $Sb_2Te_3$ being a 3D TI is 4 quintuple layers (QL). Below 4 QL, an energy gap that results from the interaction of surface states from the opposite sides of the film and thus the thickness dependence are observed.

Our experiments were carried out in a combined ultrahigh vacuum MBE-STM system (Unisoku). The base pressure of the system is better than $1\times10^{-10}$ Torr. In the MBE growth of $Sb_2Te_3$ films, highly n-doped 6H-SiC(0001) covered mainly by bilayer graphene was used as substrate [19]. The inert graphene avoids interface chemical reaction and leads to an atomically sharp interface [19], which is crucial for the current study. $Sb_2Te_3$ films were prepared by thermal evaporation of high-purity Sb (99.9999%) and Te (99.9999%) from two standard Knudsen cells. The temperatures of Sb source, Te source and the substrate were set at 330 °C, 225 °C and 230 °C, respectively, which results in a growth rate of ~ 0.2 QL/min. While the defects in $Bi_2Te_3$ can be n- or p-type by different growth conditions [20] and in $Bi_2Se_3$ can only be n-type [19], two types of intrinsic defects, Sb vacancy and $Sb_{Te}$ anti-site defect, are all proved to be p-type in $Sb_2Te_3$ [21]. The defect formation is very sensitive to the effective flux ratio between Sb and Te, which can be tuned by changing the substrate temperature because of the temperature dependent sticking coefficient of Sb and Te on the surface. The latter defect appears at relatively higher substrate temperatures, while the former dominates at lower temperatures (both can be minimized by fine-tuning of growth conditions) [21]. In this work, three kinds of samples (type-I, II, and III) were investigated. Type-I (~ 30 QL thick, Fig. 1) samples were prepared at medium substrate temperature (~ 230 °C). Type-II (3-8 QL thick, Fig. 2) contain more substitutional defects, which were prepared by annealing type-I samples at ~ 250 °C. Type-III samples (1-8 QL, Fig. 3) were grown at relatively low temperature (~ 200 °C). In this case, the dominating defects are Sb vacancies.

The morphology of type-I sample is shown in Fig. 1a, in which wide terraces separated by regular steps with a height of about 1.01 nm can be seen. The insert in Fig. 1a shows the atomic resolution image of the surface. The surface is Te-terminated (111) surface with a lattice constant of about 4.26 Å. High-density (~ $10^{12}$ cm$^{-2}$) clover-shaped defects can be clearly seen in Fig. 1b. They are Sb$_{Te1}$ and Sb$_{Te3}$ in the first QL, as identified by atomic resolution images [21]. Here Te1 and Te3 correspond to the first-layer and third-layer Te atoms from the surface.

The *dI/dV* spectrum in STM measures the local density of states (LDOS) of a sample surface at various energies. In Fig. 1c, we show *dI/dV* spectrum of type-I sample from -250 meV to +400 meV at zero magnetic field. The sharp increase in LDOS around 0 meV ($E_F$) and above 300 meV indicates a bulk energy gap of ~300 meV, where the energy of about 15 meV above $E_F$ can be attributed to the bulk valence band (VB) edge. In the bulk gap, a typical V-shaped spectrum with a conductance minimum representing the Dirac point (DP) at 100 meV above $E_F$ is seen, which is also supported by the occurrence of the zero$^{th}$ Landau level at this energy in the magnetic field (Fig. 1d). The nearly zero tunneling conductance at DP and the V-shaped LDOS are consistent with the Dirac cone structure of the surface states. Therefore, we know that the DP in the 30 QL $Sb_2Te_3$ film is separated from the bulk VB edge by 85 meV.

The 2D massless DF nature of the surface states can be revealed by Landau level (LL) spectroscopy [22-23]. In the presence of a perpendicular magnetic field, the momentum and energy of surface state electrons are quantized into discrete values, $E_n$ and $k_n = \sqrt{2e|n|B/\hbar}, n = 0, \pm 1, \pm 2, ...$ [23], where $e$ is the electric charge of an electron, $n$ is the LL index, $B$ is the magnetic field and $\hbar$ is the reduced Plank constant. The values of $E_n$ are determined by Lorenzian fits of peaks in the LL spectra. The LL spectrum of massless DFs follows the unique square root dependent sequence,

$$E_n = E_D + \text{sgn}(n) v_F \sqrt{2e\hbar |n| B}, n = 0, \pm 1, \pm 2, ..., \qquad (1)$$

where $v_F$ is the Fermi velocity, and $E_D$ is the DP energy.

Figure 1d shows the LL spectrum of type-I sample at a magnetic field of 7 T (curve). A series of non-equally spaced sharp LL peaks can be observed. One of the unique features of the LL sequence is the appearance of a peak at DP where the LDOS is basically zero in the zero-field spectrum (Fig. 1c). This $LL_0$ peak is independent of $B$ (Fig. 2b and Fig. 4b). It is an indication of the chiral nature of $Sb_2Te_3$ surface states, which is closely related to the nontrivial Berry's phase [24]. Below the zero mode, at least two peaks can be explicitly resolved, though the intensity is suppressed probably due to a coupling to the VB states. The DP of $Sb_2Te_3$ surface states is well above its VB, which is in sharp contrast to $Bi_2Te_3$ with DP buried in VB [6] and to $Bi_2Se_3$ with DP in close proximity to VB [7, 22-23]. We plot the LL energies $E_n$ at 7 T versus the square root of LL index $sgn(n)\sqrt{|n|}$ in Fig. 1d (squares). The dispersion deviates from linearity and is a convex function at DP, implying a non-zero mass of the surface state electrons even at DP. This is inconsistent with the massless DF nature of topological surface states near DP. We attribute it to the tip-gating effect [22], which is prominent when $E_F$ lies inside the bulk gap and close to DP.

The tip effect can be eliminated in highly doped and thinner $Sb_2Te_3$ films. Type-II sample was prepared for this purpose: its surface defect density is approximately 4 times of that of type-I sample (Fig. 2a). Moreover, in a-few-QL samples like type-II, the substrate and highly doped film offer a better screening to the tip's electric field, and minimize the tip-induced band-bending. Figure 2b displays the LL spectrum series of a type-II sample at different magnetic fields, where the LLs become notable at 2 T. The field-independent DP is located at ~126 meV above $E_F$, implying a higher doping level. The LL energies $E_n$ at different $B$ are plotted against $sgn(n)\sqrt{|n|B}$ related to the quantized momentum in Fig. 2c. A nearly perfect linearity characteristic of the 2D massless DFs is immediately evident. The resulted Fermi velocity $v_F$ is $4.3 \times 10^5$ ms$^{-1}$. The observation indicates that the non-linearity in Fig. 1d is not intrinsic.

We then investigate the thickness dependence of the surface states. For a TI film below a critical thickness, the coupling between top and bottom surface states with a finite decay length will open an energy gap. For $Bi_2Se_3$, the critical thickness as determined by ARPES is 6 QL [16]. Here similar 3D-2D crossover phenomenon was also observed. Type-III sample was used because terraces with a thickness of 1-2 QL only appear at relatively low temperature of the graphene substrate.

Figures 3(a-d) show evolution of bulk quantum well states (QWS) and surface states with a film thickness from 1 QL to 4 QL. The bulk gap determined by the lowest conduction band (CB) and highest VB QWS (CB1, VB1) decreases with the increasing thickness. In the gap, there is clear evidence that the surface state opens a new and smaller gap below 4 QL. This gap is about 255 meV at 2 QL and decreases to 60 meV at 3 QL. At 1 QL, the lower surface state branch is absent, the sharp increase in LDOS at -550 meV is ascribed to the VB QWS edge and no surface state is present above this energy (Fig. 3a). This situation is consistent with theoretic calculations [8]. The gap at 3 QL is manifested with absence of the zero$^{th}$ LL in the spectrum at 7 T, where LLs in both of the upper and lower surface state branches can be seen (Fig. 3c). At 4 QL, the surface state gap nearly vanishes, with the zero-conductance point at -15 meV. The uptrend of QWS and surface state energy positions with film thickness (Fig. 3e) is a result of substrate's n-doping effect on $Sb_2Te_3$ films, which decays with thickness [19].

The conductance minimum in LDOS at 4 QL in Fig. 3d is indeed the DP of surface states, as demonstrated by the LL spectroscopy (Figs. 4a-b) of a type-II sample (to avoid tip-effect). At 4 QL, similar to 7 QL, the emergence of field-independent zero$^{th}$ peak and the good linear fitting in LL peaks (Fig. 4c) at various $B$ indicate that 4 QL $Sb_2Te_3$ film is a 3D TI. The DP at 4 QL is ~ 82 meV, lower in energy than that at 7 QL because of substrate-doping. The Fermi velocity $v_F$ is $\sim 4.6 \times 10^5$ ms$^{-1}$, a little larger than that in 7 QL films.

From the high resolution data of the LL peaks in our experiment, the quasiparticle lifetime $\tau$ can be extracted. As shown in Fig. 4d, the LL peak width shows a minimum at $E_F$, duplicating the enhanced intensity of LLs around zero energy (Fig. 4b). At the energy around LL$_0$, there is another minimum in the peak-width distribution. This phenomenon might be correlated with three main possible scattering channels: electron-electron interaction, disorder, and electron-phonon coupling. In our case, electron-phonon coupling can be excluded as the main factor of LL width broadening, because it will only lead to an increase in the LL width above certain phonon energy, above which the distribution curve is flat.

Disorder in our atomically flat films mainly comes from the intrinsic substitutional defects (Sb$_{Te}$), which can also be ruled out for the following two reasons. First, disorder will broaden the LL peaks around $E_F$ [26] and narrow them at energies away from $E_F$, which is contrary to our observation. Second, the similar peak width of type-I and II samples (Fig. 4d) strongly suggests that disorder has a little effect. Therefore, Sb$_{Te}$ induced disorder doesn't have much effect on the quasiparticle lifetime. It is understandable because disorder or impurity potential may only induce intra-band (intra-surface states) scattering when the quasiparticle energy lies well within the bulk gap in our case. In topological insulators, the intra-band scattering channel is limited by the helical nature of surface states, provided that the Fermi-surface is not warped [27]. The quasiparticle has a large probability of not being scattered. Thus this observation can be a testimony of topological surface states' helical nature.

We attribute this unusual peak-width distribution to intra-band electron-electron interaction (by electron-hole pair generation). Inter-band interaction (between SS and bulk state electrons) is excluded because of the relatively large bulk gap in our case (injected electron energy is too small to generate electron-hole pair in the bulk). The decay rate of injected electron through this process increases with energy away from $E_F$. The enhanced $\tau$ at low energies has been observed and ascribed to electron-electron interactions in graphene [28-29] and $Bi_2Se_3$ [23]. However, this monotonic trend of scattering rate with respect to energy is modified by the Dirac cone shape of surface states. The electron injected into the surface states just above the Dirac energy has few relaxation channels that satisfy conservation rules, similar to that in graphene [30] which also has an energy band of Dirac cone shape. This leads to an enhanced quasiparticle lifetime at energies around the DP. Thus electron-electron interaction and the Dirac cone together account for quite well the Landau level width distribution and is suggested as the main relaxation channel for quasiparticles in $Sb_2Te_3$ thin films. Electron-electron interaction may be increased further in the presence of a magnetic field. Recently, transport measurement on $Bi_2Se_3$ films suggests the essential role played by electron-electron interaction in the transport of topological insulators [31-32].

The peak-width value of about 4 meV at two minimums leads to a quasiparticle lifetime of about 0.2 ps, yielding a mean free path of about 80 nm by multiplying it with the DF's Fermi velocity in $Sb_2Te_3$. This value is comparable to the terrace size in our $Sb_2Te_3$ films, suggesting that the step edge can act as the scattering source and give the upper-limit of mean free path in $Sb_2Te_3$ films. Recent work also supports that the in-gap bound states could be induced by the step edge [33]. Compared to the strong impurity induced in-gap resonances [34], $Sb_{Te}$ defects in our films can be defined as 'weak' impurities that will not violate the topological protection of surface states.

In summary, we have studied the topological surface states of $Sb_2Te_3$ MBE thin films using LL spectroscopy. The exotic property of TI surface states in $Sb_2Te_3$ is demonstrated by revealing the linear dispersion of surface states and the field-independent zero$^{th}$ LL at DP. The DP in $Sb_2Te_3$ is found to be well separated from the bulk states. By analyzing the LL peak-width distribution, we found that it is not impurity induced disorder but electron-electron interaction that limits the quasiparticle lifetime in $Sb_2Te_3$. We further show that the 3D-to-2D crossover of the surface states in $Sb_2Te_3$ occurs at 4 QL.

**Acknowledgements:** This work was supported by National Science Foundation and Ministry of Science and Technology of China. All STM topographic images were processed by WSxM software (www.nanotec.es).

# References

[1]  M. Z. Hasan and C. L. Kane, Rev. Mod. Phys. **82**, 3045 (2010).
[2]  X. -L. Qi and S.-C. Zhang, Rev. Mod. Phys. **83**, 1057 (2011).
[3]  C. L. Kane and E. J. Mele, Phys. Rev. Lett. **95**, 146802 (2005).
[4]  L. Fu and C. L. Kane, Phys. Rev. B **76**, 045302 (2007).
[5]  H. J. Zhang *et al.*, Nat. Phys. **5**, 438 (2009).
[6]  Y. L. Chen *et al.*, Science **325**, 178 (2009).
[7]  Y. Xia *et al.*, Nat. Phys. **5**, 398 (2009).
[8]  G. Wang *et al.*, Nano Res. **3**, 874 (2010).
[9]  Z. Alpichshev *et al.*, Phys. Rev. Lett. **104**, 016401 (2010).
[10] P. Roushan *et al.*, Nature **460**, 1106 (2009).
[11] J. Seo *et al.*, Nature **466**, 343 (2010).
[12] T. Zhang *et al.*, Phys. Rev. Lett. **103**, 266803 (2009).
[13] J. Chen *et al.*, Phys. Rev. Lett. **105**, 176602 (2010).
[14] C.-X. Liu *et al.*, Phys. Rev. B **81**, 041307 (2010).
[15] H.-Z. Lu *et al.*, Phys. Rev. B **81**, 115407 (2010).
[16] Y. Zhang *et al.*, Nat. Phys. **6**, 584 (2010).
[17] M. Konig *et al.*, J. Phys. Soc. Jpn. **77,** 031007 (2008).
[18] D. Hsieh *et al.*, Phys. Rev. Lett. **103,** 146401 (2009).
[19] C. L. Song *et al.*, Appl. Phys. Lett. **97**, 143118 (2010).
[20] G. A. Wang *et al.*, Adv. Mater. 23, 2929 (2011).
[21] Y. P. Jiang *et al.*, arXiv:1111.0817v1 (2011).
[22] P. Cheng *et al.*, Phys. Rev. Lett. **105**, 076801 (2010).
[23] T. Hanaguri *et al.*, Phys. Rev. B **82**, 081305 (2010).
[24] Y. Zhang *et al.*, Nature **438**, 201 (2005).
[25] For 3 QL films, the Sb vacancies degrade the spectrum quality in type-III sample, while the substitutional defects do not in type-I and II samples, probably because Sb vacancies may have in-gap resonances.
[26] N. M. R. Peres, F. Guinea, and A. H. Castro Neto, Phys. Rev. B **73**, 125411 (2006).
[27] L. Fu, Phys. Rev. Lett. **103**, 266801 (2009).
[28] G. Li, A. Luican, and E. Y. Andrei, Phys. Rev. Lett. **102**, 176804 (2009).
[29] Gonz *et al.*, Phys. Rev. Lett. **77**, 3589 (1996).
[30] A. Bostwick *et al.*, Nat. Phys. **3**, 36 (2007).
[31] M. H. Liu *et al.*, Phys. Rev. B **83**, 165440 (2011).
[32] J. Wang *et al.*, Phys. Rev. B **83**, 245438 (2011).
[33] Z. Alpichshev *et al.*, Phys. Rev. B **84**, 041104(R) (2011).
[34] A. M. Black-Schaffer and A. V. Balatsky, arXiv:1110.5149v1 (2011).

**Figure captions:**

FIG. 1 (color online). (a) STM topographic image of 30 QL type-I sample prepared by MBE. Tunneling conditions: $V_{bias}$ = 5.0 V, I = 50 pA. The bias voltage is applied to the sample. The insert shows an atomic resolution image of $Sb_2Te_3$ (111) surface ($V_{bias}$ = 20 mV, I = 0.1 nA). The brighter spots correspond to Te atoms. (b) STM topographic image ($V_{bias}$ = 1.0 V, I = 50 pA) showing the typical surface defects of the sample. (c) Bulk ($V_{bias}$ = 0.3 V, I = 50 pA) and surface state (SS) spectrum ($V_{bias}$ = 0.25 V, I = 0.2 nA) in the energy gap. Because of the much lower SS DOS compared with bulk DOS, a much closer tip-sample tunneling gap is needed to reveal the detail of SS. (d) Landau quantization of topological surface states at 7 T (curve), relation between LL energies and $\text{sgn}(n)\sqrt{|n|}$ (squares), and a linear fitting to the relation (line). The LL peak positions were determined by multi-peak Lorentzian fits. The bias modulation was 1 mV at 987.5 Hz.

FIG. 2 (color online). (a) STM topographic image ($V_{bias}$ = 1.0 V, I = 50 pA) showing the surface defects of 7 QL type-II sample. (b) $dI/dV$ spectra ($V_{bias}$ = 0.25 V, I = 0.2 nA) from 0 T to 7 T. The dotted line indicates a field-independent $n = 0$ LL at Dirac point. Spectra under different fields are shifted vertically for clarity. (c) LL energies under various magnetic fields plotted versus $\text{sgn}(n)\sqrt{|n|B}$. The line is a linear fitting to the data.

FIG. 3 (color online). (a-d) $dI/dV$ spectra of 1-4 QL Type-III $Sb_2Te_3$ films. The bulk QWS are indicated by arrows. The corresponding surface state spectrum for each thickness is displayed in the lower panel. The insert in (a) represents two sets of surface states on the opposite surfaces. For 3 QL films, the surface state spectrum at 7 T is also presented together with zero-field spectrum in the lower panel of (c). They were taken on type-II samples [25]. The spectra are shifted vertically for clarity. (e) The thickness-dependent QWS, lowest unoccupied and highest occupied surface state branches (LUS and HOS). The positions of LUS and HOS are indicated by arrows in the surface state spectra.

FIG. 4 (color online). (a-b) *dI/dV* spectra ($V_{bias}$ = 0.2 V, I = 0.2 nA) of a 4 QL type-II sample without magnetic field (a) and under perpendicular magnetic field (b). (c) The fitting of the LL energies versus $\text{sgn}(n)\sqrt{|n|B}$ for magnetic fields from 2 to 7 T. The line is a linear fitting to the data. (d) Full width at half maximum of LL peaks at different energies in (b) and Fig. 1(d) fitted by Lorentzian function. The hollow squares correspond to the LL peaks in type-I sample in Fig. 1(d) and have been shifted by the energy difference between DPs of the two different types of samples.

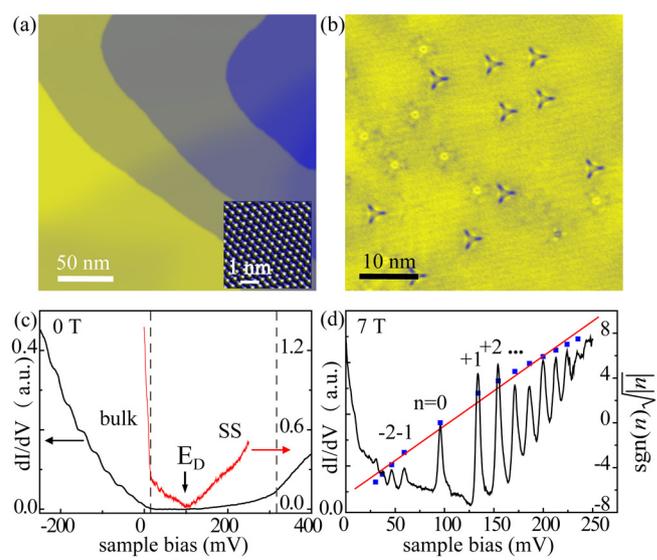

**Figure 1**

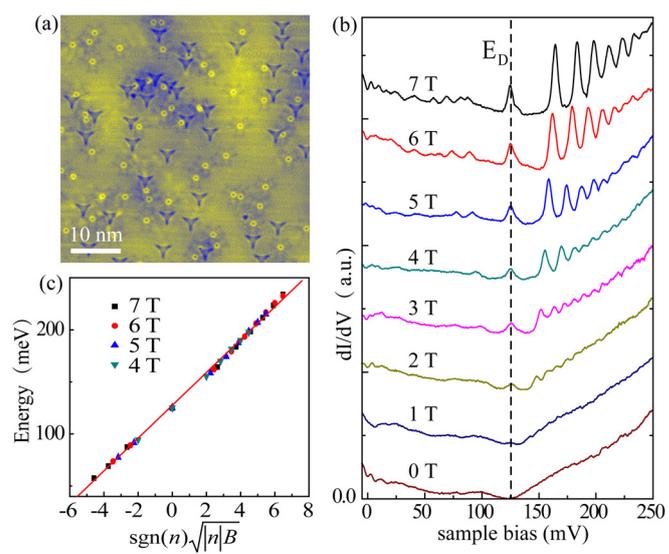

**Figure 2**

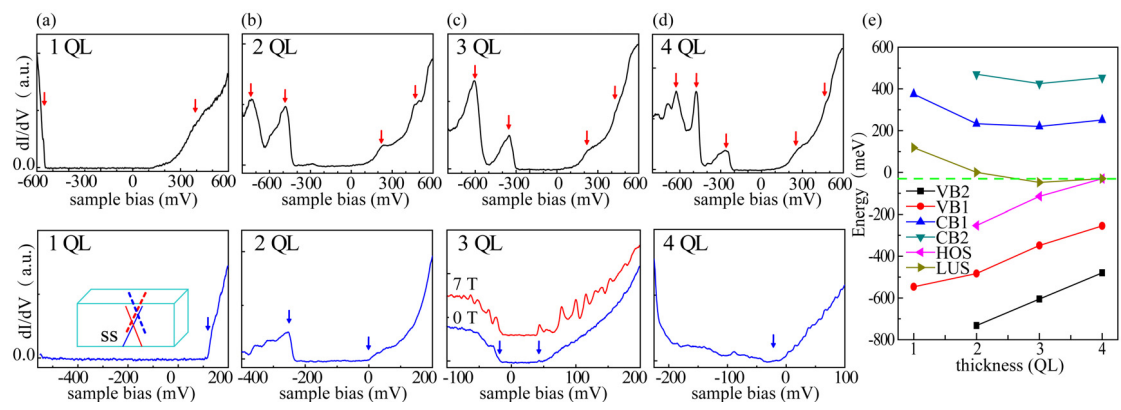

**Figure 3**

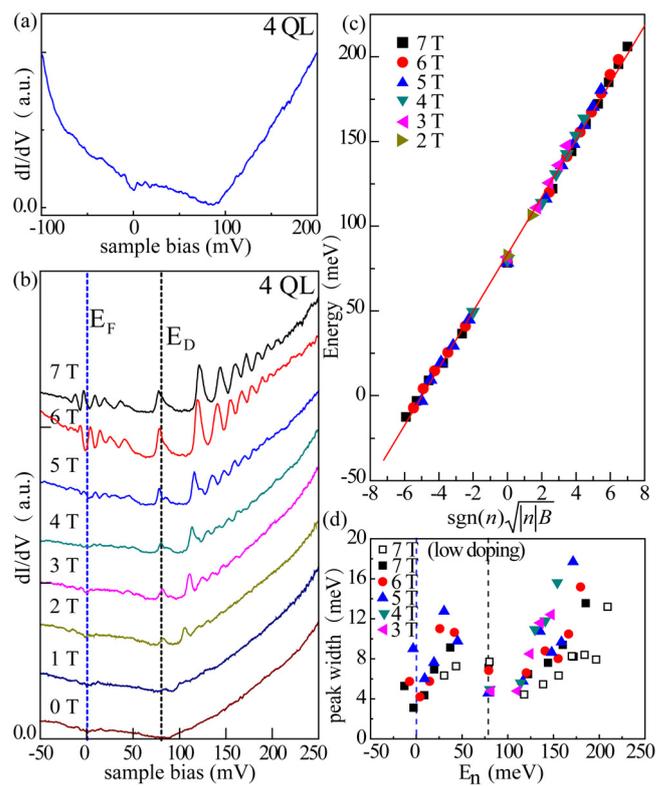

**Figure 4**